\documentclass[twocolumn]{aastex62}

\usepackage{xspace}

\newcommand{\FdTz}{$\mathtt{F}_{\dot{E}}^{\delta} $\xspace}
\newcommand{\FsfTz}{$\mathtt{F}_{\left < a \right >}^{\delta}$\xspace}
\newcommand{\FsfTq}{$\mathtt{F}^\mathtt{1/4T}_{\left < a \right >}$\xspace}
\newcommand{\FsfTs}{$\mathtt{F}^\mathtt{1/16T}_{\left < a \right >}$\xspace}
\newcommand{\FsfTo}{$\mathtt{F}^\mathtt{1T}_{\left < a \right >}$\xspace}
\newcommand{\FdTzSup}{$\mathtt{F}_{\dot{E}}^{\delta}\mathtt{M2}$\xspace}
\newcommand{\FsfTzSup}{$\mathtt{F}_{\left < a \right >}^{\delta}\mathtt{M2}$\xspace}
\newcommand{\FsfToSup}{$\mathtt{F}^\mathtt{1T}_{\left < a \right >}\mathtt{M2}$\xspace}
\newcommand{\RMSa}{$\left < a \right >$\xspace}
\newcommand{\Tcorr}{$T_{\mathrm{corr}}$\xspace}

\newcommand{\V}[1]{\mathbf{#1}}
\newcommand{\bra}[1]{\left ( #1 \right )}

\newcommand{\Ms}{\mathrm{M_s}}
\newcommand{\Ma}{\mathrm{M_a}}
\newcommand{\pd}{\partial}

\newcommand{\pth}{p_\mathrm{th}}

\newcommand{\eqref}[1]{(\ref{#1})}

\begin{document}
\title{As a matter of force  -- Systematic biases in idealized turbulence simulations}
\shorttitle{Biases in turbulence simulations}
\shortauthors{Grete, O'Shea \& Beckwith}
\correspondingauthor{Philipp Grete}
\email{grete@pa.msu.edu.}

\author[0000-0003-3555-9886]{Philipp Grete}
\affiliation{
Department of Physics and Astronomy,
Michigan State University, East Lansing, MI 48824, USA}
\author{Brian W. O'Shea}
\affiliation{
Department of Physics and Astronomy,
Michigan State University, East Lansing, MI 48824, USA}
\affiliation{
Department of Computational Mathematics, Science and Engineering,
Michigan State University, East Lansing, MI 48824, USA}
\affiliation{
National Superconducting Cyclotron Laboratory,
Michigan State University, East Lansing, MI 48824, USA}

\author{Kris Beckwith}
\affiliation{
Sandia National Laboratories, Albuquerque, NM 87185-1189, USA
}

\reportnum{SAND Number: SAND2018-4724 J}

\begin{abstract}

Many astrophysical systems encompass very large dynamical ranges in space and time, 
which are not accessible by direct numerical simulations.
Thus, idealized subvolumes are often used to study small-scale effects including the 
dynamics of turbulence.
These turbulent boxes require an artificial driving in order to mimic energy injection 
from large-scale processes.
In this Letter, we show and quantify how the autocorrelation time of the driving and its 
normalization systematically change properties 
of an isothermal compressible magnetohydrodynamic flow in the sub- and supersonic regime
and affect
astrophysical observations such as Faraday rotation.
For example, we find that $\delta$-in-time forcing with a constant energy injection leads to a steeper
slope in kinetic energy spectrum and less efficient small-scale dynamo action.
In general, we show that shorter autocorrelation times require more power in the acceleration field,
which results in more power in compressive modes that weaken the
anticorrelation between density and magnetic field strength.
Thus, derived observables, such as the line-of-sight magnetic field from rotation measures, are 
systematically biased by the driving mechanism.
We argue that $\delta$-in-time forcing is unrealistic and numerically unresolved, 
and conclude that special care needs to be taken in interpreting observational results 
based on the use of idealized simulations.

\end{abstract}

\section{Introduction}

Many astrophysical systems are governed by compressible magnetohydrodynamic (MHD)
dynamics on macroscopic scales \citep{Galtier2016}.
Moreover, given the large scales involved astrophysical systems are often in a turbulent state
\citep{Brandenburg2013}.
Compressible MHD turbulence itself is expected to be a major factor in many
processes such as magnetic field amplification via the turbulent dynamo
\citep{Brandenburg2005,Tobias2013} and particle acceleration in shock fronts
which can eventually be observed as cosmic rays \citep{Brunetti2015}.

Astrophysical observations of distant systems are typically a two dimensional map
that measure quantities along the third, integrated dimension.
In order to interpret those observations numerical simulations are often used as they
provide detailed data in four (3 spatial and a temporal) dimensions 
\citep{Burkhart2012,Burkhart2017,Orkisz2017}.
However, the large dynamical range in space and time often prohibit direct numerical
simulation of an entire system.
Thus, idealized subvolumes are used to study specific effects and processes including
small-scale turbulent dynamics in so called turbulent boxes.
Turbulent boxes are the workhorse in both astrophysical and general turbulence research, and 
are used to study a variety of aspects of turbulence including energy transfers
in the hydrodynamic (HD) \citep{Kida1990} and MHD \citep{Yang2016,Grete2017a} case, 
or HD \citep{Clark1979,Germano1991} and MHD \citep{Chernyshov2012,Grete2016a} turbulence models.
To reach a state of turbulence in simulations a large-scale driving field\footnote{
In this Letter, we use forcing, driving and acceleration interchangeably to describe 
a mechanical energy injection process.}
is commonly used.

One fundamental assumption in these simulations is that the dynamics on the large scales 
(where energy in injected)
are decoupled from the dynamics on smaller scales.
In other words, large-scale features are lost in the energy cascade towards small scales.
We show that this assumption is wrong even for simple, purely solenoidal driving.

Many different driving schemes are used in numerical turbulence research.
Here, we focus on schemes that are popular in the astrophysical turbulence community and 
often correspond to the default turbulence in a box setup in many codes.
These schemes can be differentiated by two main properties: the autocorrelation time 
and the normalization applied to the driving field.
The autocorrelation time determines on which timescale the driving field evolves.
Most commonly two extreme cases are used.
On the one hand, a $\delta$-in-time forcing calculates a new random driving field on
each timestep that is completely uncorrelated to the driving field of the previous timestep.
On the other hand, smoothly evolving driving fields are used with a given autocorrelation time
(often set to the dynamical time of the simulation) realized, e.g., 
by a stochastic process \citep{Eswaran1988}.
In both cases the driving field undergoes a random change on each timestep, which leads
to a random change in its power.
Hence, the second differentiating property is how the amplitude of the acceleration field
is normalized on each timestep.
Again, two possibilities are commonly used.
First, the driving field is normalized to have constant power over time, i.e., the 
root mean square (RMS) value \RMSa is constant.
Second, the driving field is normalized to have a constant energy injection rate $\dot{E}$ 
throughout the simulation.
Examples for $\delta$-in-time forcing with constant energy injection include
\citet{Stone1998}, \citet{Lemaster2009}, or
\citet{Kim2005}. 
Examples for $\delta$-in-time forcing with constant power include
\citet{Brandenburg2002}.
Examples using a driving field
that evolves on a dynamical timescale with constant power
include \citet{Cho2009}, 
\citet{Federrath2008},  and \citet{Schmidt2009}.

A previous study by \citet{Yoon2016} analyzed simulations with two different driving
mechanisms: a $\delta$-in-time forcing normalized to $\dot{E}$, and 
a driving field with a finite correlation time and constant power.
They find differences in the correlation between density and magnetic field strength and in
statistical moments of density related fields including the probability density function, the
dispersion measure and Faraday rotation measure.
\citet{Yoon2016} attribute those differences to a link between the autocorrelation time of the 
driving and the ability of the system to reach pressure equilibrium.
Here, we go one step further and show that the autocorrelation time is only a secondary
parameter.
The primary driver in the observed differences is the power in the compressive modes of the 
resulting flow.

In particular, we show that the power in the acceleration field, which varies with the autocorrelation time
for similar stationary regimes, directly affects compressive modes in the simulation.
Varying compressive power, in turn, results in different statistical properties such
as the slope in the kinetic energy spectrum or the correlation between density and magnetic
field strength.
Moreover, these differences manifest in changing observable quantities, 
e.g., Faraday rotation measures.
Thus, a systematic bias is introduced when simulations are used as a basis 
to interpret observations such as the line-of-sight magnetic field strength.
The results presented cover the sub- and (mildly) supersonic regime.
Thus, they are
particularly relevant for turbulence in the warm ionized medium
\citep{Iacobelli2014,Herron2016}.

This Letter is organized as follows.
In Section~\ref{sec:method}, we introduce the simulations and the implementation of 
the different driving mechanisms.
In Section~\ref{sec:results}, we present the key results and differences in the simulations, 
and discuss the implications in Section~\ref{sec:disc}.
Finally, we conclude in Section~\ref{sec:conclusions} and provide future directions.

\section{Method}
\label{sec:method}
In this work we are dealing with the compressible, ideal MHD
equations
\begin{eqnarray}
\label{eq:rho}
\pd_t \rho +  \nabla \cdot \bra{\rho \V{u}} &= 0, \\
\label{eq:rhoU}
\pd_t \rho \V{u} + \nabla \cdot \bra{ \rho \V{u} \otimes \V{u} -  \V{B} \otimes \V{B}}
+ \nabla \bra{ \pth +B^2 /2} &= \rho \V{a}, \\
\label{eq:B}
\pd_t \V{B} -  \nabla \times \bra{ \V{u} \times \V{B}}  &= 0,
\end{eqnarray}
that are closed by an isothermal equation of state.
The symbols have their usual meaning, i.e., density~$\rho$, velocity~$\V{u}$,
thermal pressure~$\pth$, and magnetic field~$\V{B}$, which includes a factor~$1/\sqrt{4\pi}$.
Vector quantities that are not in boldface refer to the $L^2$ norm of the vector and
$\otimes$ denotes the outer product.
The details of the acceleration field~$\V{a}$ that we use to mechanically drive our 
simulations are described below in Section~\ref{sec:forcing}.

\subsection{Simulations}
We use a modified version\footnote{
  The implementation of the stochastic forcing used in this paper is available at
\url{https://github.com/pgrete/Athena-Cversion}.}
of the astrophysical MHD code \textsc{Athena 4.2} \citep{Stone2008}.
All simulations use second order reconstruction with slope-limiting 
in the primitive variables, the HLLD Riemann solver, constrained transport for the 
magnetic field, and the MUSCL-Hancock integrator \citep{Stone2009} on 
a uniform, static grid with $512^3$ cells.
We start with uniform initial conditions (all in code units) $\rho = 1$, $\V{u} = \V{0}$, and 
$\V{B}~=~\bra{1/6,0,0}^{\mathrm{T}}$ (subsonic) or 
$\V{B}~=~\bra{2/3,0,0}^{\mathrm{T}}$ (supersonic) corresponding to initial
plasma betas of 72 and 4.5, respectively.
The two different initial $\V{B}$ lead to the the same
Alfv\'enic Mach number in both regimes. We evolve the
system for five dynamical times $\mathrm{T} = \mathrm{V}/0.5\mathrm{L}$.
Here, $\mathrm{V}$ is the characteristic velocity in the stationary phase, 
which corresponds to the root mean square (RMS) sonic Mach number $\Ms$ 
as we fix the isothermal sound speed to~1.
In the subsonic simulations $\mathrm{V} = 0.5$ and in the supersonic simulations
$\mathrm{V} = 2$.
The characteristic length $0.5$L is half the box size ($\mathrm{L}=1$), because our acceleration
spectrum is parabolic \citep{Schmidt2009} and peaks at $k = 2$ 
(using normalized wavenumbers).
The acceleration field is purely solenoidal, i.e., $\nabla \cdot \V{a} = 0$.
We store 20 equidistant snapshots per dynamical time.
The stationary phase is reached after approximately $2.5$T, and we calculate statistical
properties for this phase based on 50 snapshots between $2.5\mathrm{T} < t < 5\mathrm{T}$.

\subsection{Forcing mechanisms}
\label{sec:forcing}

In order to quantify the influence of the autocorrelation time of the driving field
we implemented a stochastic forcing mechanism as presented by~\citet{Schmidt2009}
in \textsc{Athena}, and conduct four identical simulation in the subsonic regime
that vary only in their autocorrelation time (and their RMS acceleration value \RMSa as explained in 
subsection~\ref{sec:acc}). 
We identify these simulations with
\FsfTo, \FsfTq, \FsfTs, and \FsfTz~corresponding to correlation times of 
$1$T, $0.25$T, $0.0625$T, and $10^{-9}$T, and \RMSa of $1$, $\sqrt{2}$, $2$ and $40$, respectively.
The last simulation is effectively $\delta$-in-time correlated as the smallest timestep
in the simulation is~$>10^{-5}$.

This allows a direct comparison to the existing forcing implementation in \textsc{Athena}.
It produces $\delta$-correlated realizations that are normalized by the energy input rate.
We conduct one simulation in the subsonic regime  with this mechanism and set 
$\dot{E} = 0.1$ in order to reach the
same RMS sonic Mach number as the other simulations.
We refer to this simulation as \FdTz throughout the paper.

To verify our findings in a more compressive regime 
we conduct supersonic simulations of the three corner cases: $\delta$-in-time normalized to
$\dot{E}$~$(= 6.4)$, $\delta$-in-time normalized to \RMSa~$(=1000)$, 
and 1T-correlated normalized to \RMSa~$(=16)$.
These simulations are referred to as \FdTzSup, \FsfTzSup, and \FsfToSup, respectively, and 
are separately presented in Section~\ref{sec:supersonic}.

\section{Results}
\label{sec:results}

\subsection{Temporal evolution}
\begin{figure}[htbp]
\centering
\includegraphics{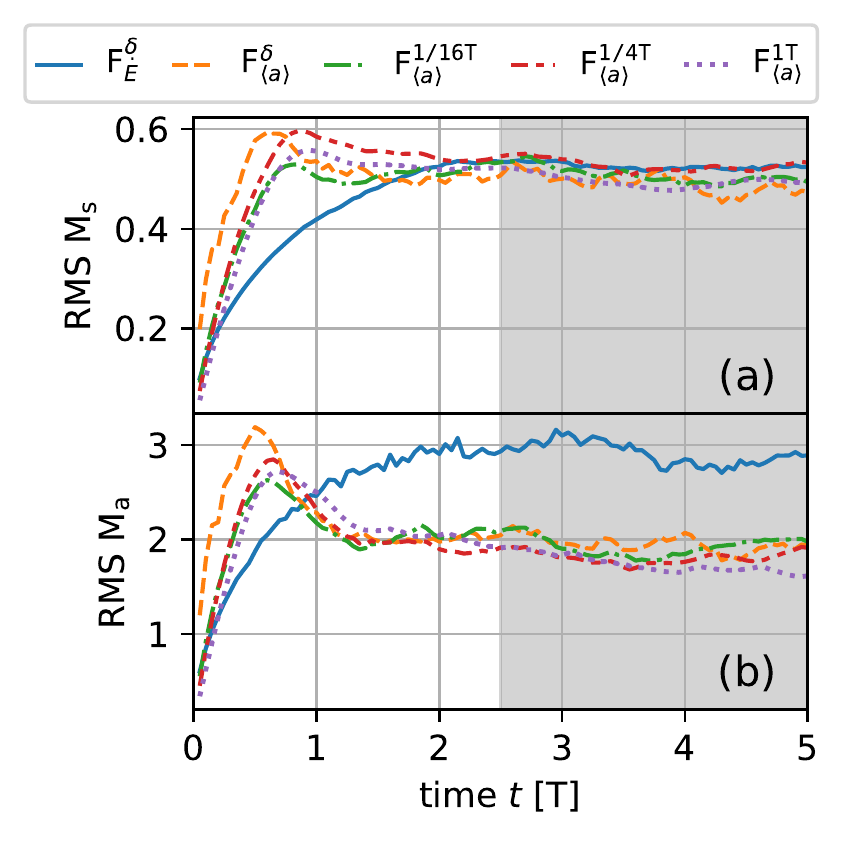}
\caption{Temporal evolution of the spatial root mean square (RMS) 
sonic Mach number (a) and  Alfv\'enic Mach number (b).
The gray area between $2.5\mathrm{T} \leq t \leq 5\mathrm{T}$ indicates the temporal range we use
as stationary regime throughout the paper.
}
\label{fig:time-evol}
\end{figure}
All subsonic simulations reach a stationary regime with 
a sonic Mach number $\Ms \approx 0.5$ after $\approx 2.5$T, as
illustrated in Fig.~\ref{fig:time-evol}(a).
In contrast to this, the Alfv\'enic Mach numbers $\Ma = \sqrt{\rho}u/B$ 
varies substantially between different normalizations.
While \FdTz is $\approx3$ in the stationary regime, it reaches only $\approx 2$ 
for all other simulations, shown in Fig.~\ref{fig:time-evol}(b).
Given that $\Ma$ is a proxy for the ratio of kinetic to magnetic energy, 
a higher value for \FdTz corresponds to a lower saturation value of the magnetic
energy.

\begin{figure}[htbp]
\centering
\includegraphics{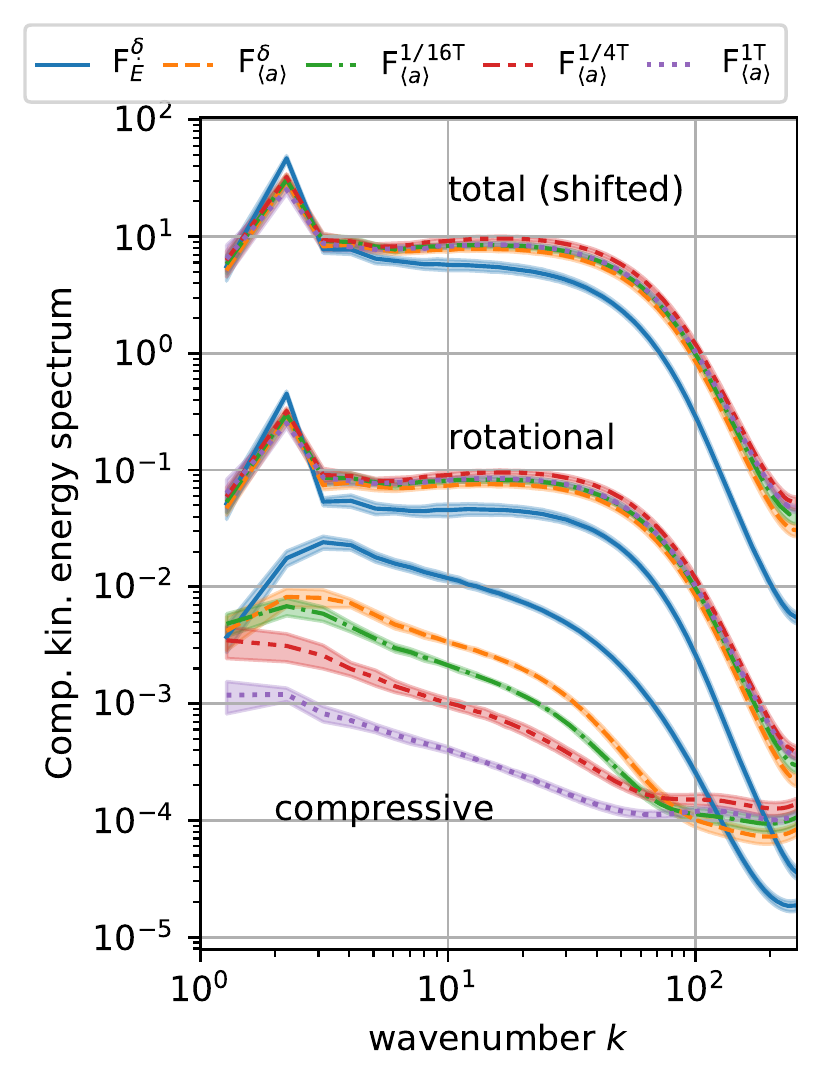}
\caption{Compressive ($\V{u}_\mathrm{c}$), rotational $\V{u}_\mathrm{s}$,
and total (i.e., compressive plus rotational)
kinetic energy spectra compensated by $k^{4/3}$.
The total energy  spectra are shifted
vertically by a factor of 100 for clarity.
All lines correspond to the temporal mean during the stationary phase and the shaded areas
indicate the standard deviation over time.
The energy spectra are calculated based on the Fourier transforms of $\sqrt{\rho}\V{u}$
and are virtually identical to the ones based on $\V{u}$.
}
\label{fig:KinEnSpec}
\end{figure}
We attribute this to a less efficient small-scale dynamo, which is driven
by rotational motion.
Figure~\ref{fig:KinEnSpec} shows the mean kinetic energy spectra 
(rotational, compressive, and total) in the stationary regime based on 
the Helmholtz decomposition of the velocity field 
$\V{u}~=~\V{u}_\mathrm{c} + \V{u}_\mathrm{s} + \V{u}_0$ with
$\nabla \cdot \V{u}_\mathrm{s} = 0$ and $\nabla \times \V{u}_\mathrm{c} = \V{0}$.
The simulation \FdTz has significantly less power ($\approx 50$\%
compared to the other simulations) 
in rotational modes on scales smaller than the injection scale.
Similarly, there is much more power in compressive motions in that
simulation, to the degree that the total (compressive plus rotational) 
kinetic energy spectrum
exhibits a different slope in the power law regime $6 \lesssim k \lesssim 20$.
It is steeper as expected from a strongly compressive turbulence
phenomenology \citep{Burgers1948}.

The rotational and total energy spectra for the simulations
\FsfTz, \FsfTs, \FsfTq, and \FsfTo are all virtually identical 
given that the compressive modes are weaker by at least one order of magnitude.
However, there are clear differences in the compressive power.
With decreasing correlation time (and, thus, increasing \RMSa), there is
more power in the compressive modes.
This can be explained by a more detailed analysis of the link between 
\RMSa and \Tcorr in the following subsection.

\subsection{Linking \Tcorr, \RMSa and compressive modes}
\label{sec:acc}
In all simulations, the driving field is purely solenoidal, i.e., it carries no
compressive power itself.
Nevertheless, a strong (high \RMSa) driving is expected to seed compressive modes,
c.f., the canonical idea of wave steepening.
\begin{figure}[htbp]
\centering
\includegraphics{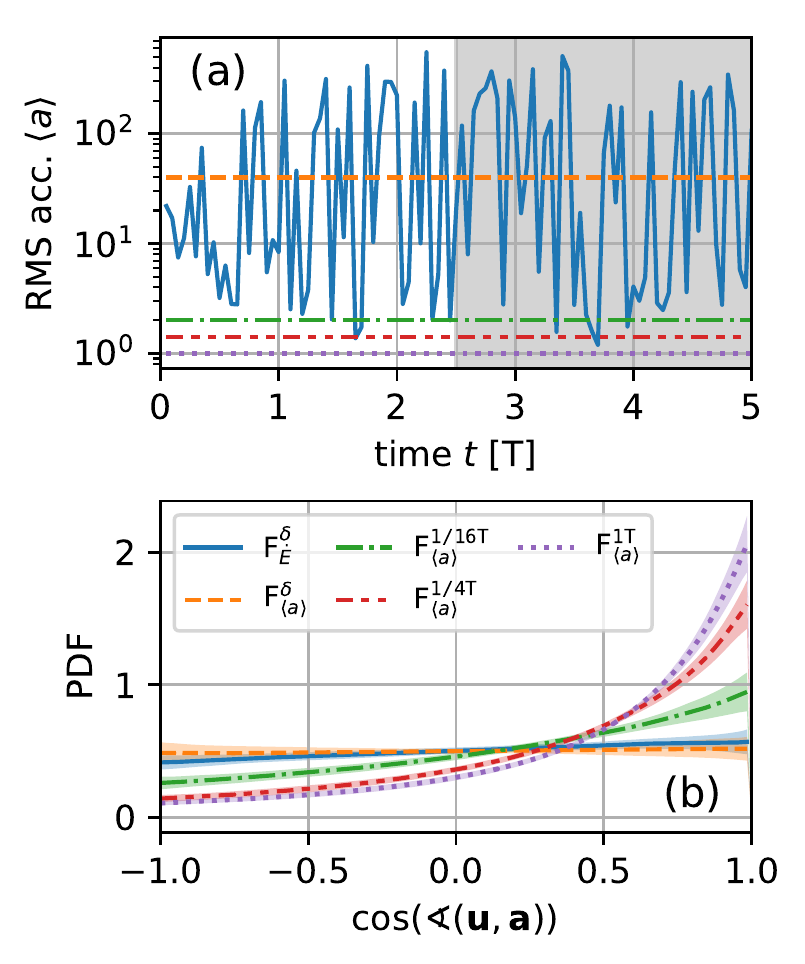}
\caption{Temporal evolution of the spatial root mean square (RMS) 
acceleration (a), and the temporal mean
PDF of the cosine of the angle between the velocity/momentum field
and the acceleration field (b).
All lines in panel (b) correspond to the temporal mean during the stationary phase (gray area) 
and the shaded colored areas indicate the standard deviation over time.}
\label{fig:acc-angle}
\end{figure}

Figure~\ref{fig:acc-angle}(a) shows \RMSa for all subsonic simulations. 
By construction \RMSa is constant for \FsfTz, \FsfTs, \FsfTq, and \FsfTo, while
it varies strongly over more than two orders of magnitude for \FdTz, which is normalized
for a constant energy injection rate~$\dot{E}$.
To keep $\dot{E}$ at a given value \RMSa can take arbitrarily large (positive) values
for the following reason.
The driving field consists of new (large-scale) random vectors at every single timestep, 
making it possible that it is locally perpendicular to the large-scale velocity field in the
entire box.
In this case, the resulting energy injection (based on the scalar product of $\V{a}$ 
and $\V{u}$) would be negligible for ``small'' \RMSa. 
Thus, during the normalization step \RMSa is increased (decreased) to match a desired $\dot{E}$
for an predominantly perpendicular (aligned) combination of flow configuration and acceleration field.

The same mechanism, i.e., the alignment of $\V{a}$ and $\V{u}$, is responsible for 
requiring a higher \RMSa value for smaller \Tcorr to reach the same $\Ms$ in the other simulations.
Figure~\ref{fig:acc-angle}(b) illustrates the mean probability density function (PDF) 
of the angle between $\V{a}$ and $\V{u}$.
A larger correlation time results in a distribution for which there is a tendency of $\V{a}$ 
and $\V{u}$ being more aligned.
This illustrates how large-scale forcing patterns, which evolve (or exist)
for a reasonable fraction of a dynamical time, leave an imprint on the large-scale
flow pattern.
Despite its clear signal in the PDFs, this alignment should not be overrated, because 
for our chosen binning a perfect alignment in the entire box would correspond to a 
$\delta$-peak with a value of 64.
Nevertheless, it is enough to influence the energy injection efficiency (via the local
scalar product between $\V{a}$ and $\V{u}$) and requiring larger \RMSa for lower \Tcorr
\citep{Eswaran1988}.
For the extreme cases, \FdTz and \FsfTz, there is no imprint of the large-scale
pattern on the flow, which is demonstrated by flat PDFs as expected.
 
While the large-scale imprint vanishes for smaller \Tcorr, another feature is introduced to the flow
by larger \RMSa: compressive modes.
At locations where $\V{u}$ and $\V{a}$ are aligned, which is always the case given the 
non-zero PDF around $\cos\bra{\sphericalangle\bra{\V{u},\V{a}}} = 1$ in Fig.~\ref{fig:acc-angle}(b),
large \RMSa lead to a strong acceleration, resulting in immediate downstream compression.
A careful examination of the compressive power spectra in Fig.~\ref{fig:KinEnSpec}
reveals signatures of this effect.
The compressive power of \FdTz peaks at $k\approx3$, which corresponds to the
smallest scales in the power spectrum of the acceleration field.
Additional signatures of this compression effect are also visible in statistics of 
the density field, as illustrated in the following subsection.

\subsection{Density field dynamics}
\begin{figure*}[htbp]
\centering
\includegraphics{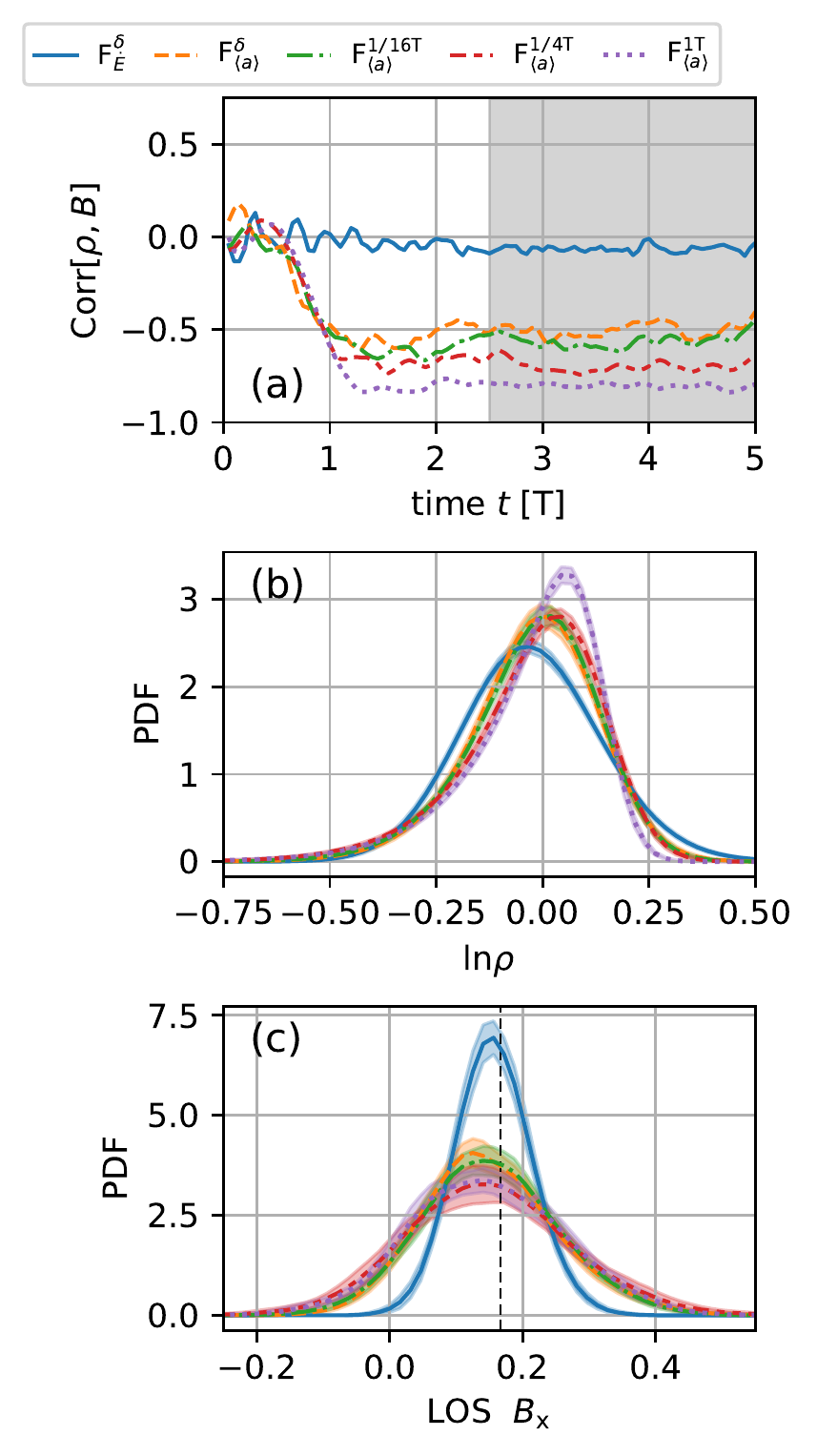}
\includegraphics{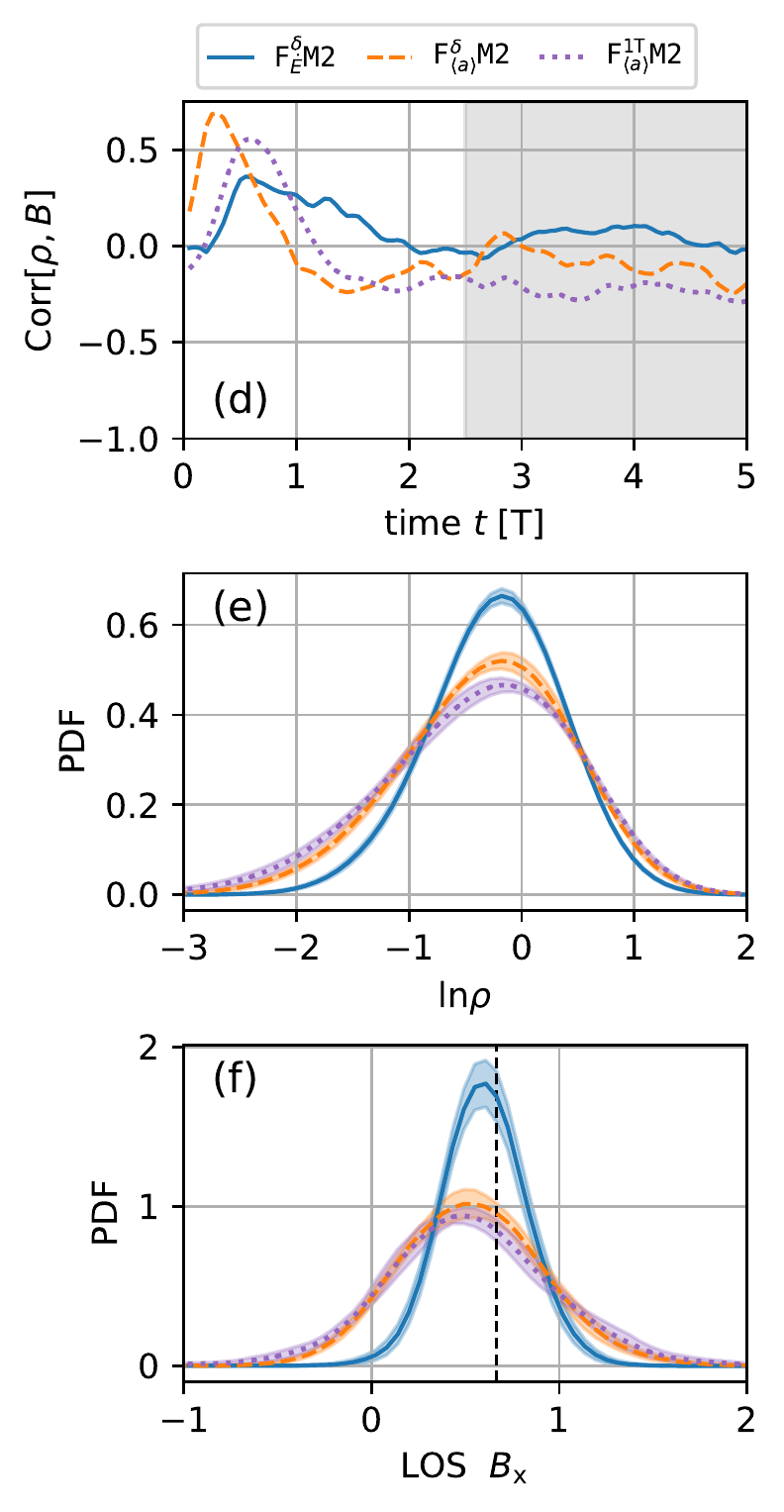}
\caption{Temporal evolution of the 
correlation between density $\rho$ and magnetic field magnitude (a,d), and
temporal mean PDFs of the logarithmic density (b,e) and line of sight magnetic field (c,f).
Subsonic simulations are shown in the left panels and supersonic simulations in the panels on the
right.
All lines in the bottom two rows correspond to the temporal mean during the stationary phase 
(gray area) and the shaded colored areas indicate the standard deviation over time.
The true line-of-sight magnetic field strength is illustrated by the vertical dashed line in
the two bottom panels.
}
\label{fig:obs}
\end{figure*}

A link between the Pearson correlation coefficient between density and magnetic field strength,
$\mathrm{Corr}\left [ \rho, B \right ]$, and the correlation time of the forcing 
has already been recognized by \citet{Yoon2016}.
However, the correlation time is only one of two integral parts, see Fig.~\ref{fig:obs}(a),
which shows the correlation coefficient over time for all subsonic simulations.
\FsfTo, \FsfTq, \FsfTs, and \FsfTz build one family for which the correlation coefficient
in the stationary regime 
increases from $-0.803(12)$, $-0.70(3)$, $-0.56(3)$, to $-0.50(4)$, respectively. 
In other words, \textit{there exists a strong anticorrelation} for a \Tcorr$ = 1$T that decreases 
with smaller \Tcorr towards to a still significant, non zero value of $-0.5$ for 
$\delta$-in-time forcing (\FsfTz).
In contrast, \FdTz exhibits virtually no correlation between the density field and magnetic
field strength $\mathrm{Corr}\left [ \rho, B \right ] = -0.064(18)$.
This is in agreement with the results of \citet{Yoon2016}, who analyzed two forcing 
configurations corresponding to our \FsfTo and \FdTz cases.
Given our additional simulations with varying \Tcorr, we argue 
that the power in the acceleration field \RMSa and not \Tcorr is
the primary driver of changes in $\mathrm{Corr}\left [ \rho, B \right ]$.
Otherwise, \FsfTz and \FdTz should yield identical results, which is not observed here.
Nevertheless, \RMSa and \Tcorr are tightly linked as shown in the 
previous subsection.

Clear differences between the simulations are also observed in the logarithmic density 
PDFs, as illustrated in Fig.~\ref{fig:obs}(b).
\FsfTo exhibits a pronounced negative skew, i.e., the low density tail is longer so that
lower than average density values are more likely than higher values.
While the skewness decreases with decreasing \Tcorr it it still present in the
\FsfTz simulation.
Again, \FdTz contrasts with these results showing an almost symmetrical distribution.
In general, the high density tails (with \FdTz allowing for the most and \FsfTo for the 
least extreme value) nicely illustrate how larger \RMSa lead to immediate compression.

Finally, in order to illustrate how these differences translate to biases in interpretations
of  astrophysical observations, we calculate line-of-sight (LOS) magnetic field strength derived
from rotation measures\footnote{
Technically, the relation is valid for the number density of thermal electrons, but given
the isothermal single fluid  MHD approximation employed we use the fluid density 
$\rho$ instead.
} as
\begin{eqnarray}
\label{eq:LOSB}
\mathrm{LOS} B_i = \frac{\int_L \rho\bra{l} B_\parallel \bra{l} \mathrm{d} l }
{\int_L \rho\bra{l} \mathrm{d} l}
\end{eqnarray}
with $B_\parallel$ being the line-of-sight component of the magnetic field.
\citet{Beck2003} derived how an (anti)correlation between $\rho$ and 
$B$ changes this measurement of the mean magnetic field strength, which is exact
for uncorrelated fields.
For anticorrelated fields~Eq.~\eqref{eq:LOSB} underestimates the true LOS~$B$.
This effect can be observed in Fig.~\ref{fig:obs}(c) where the mean PDFs of
the LOS~$B_x$ (over the $512^2$ available lines-of-sight) are shown.
With increasing anticorrelation from \FdTz, \FsfTz, \FsfTs, \FsfTq, to \FsfTo
the derived values of $0.1640(8)$, $0.1589(12)$, $0.1584(12)$, $0.1558(8)$, and 
$0.1533(11)$, respectively, deviate further from the real value $0.1667$. 
More strikingly, the PDF of \FdTz is much more peaked with a
standard deviation of $0.058(3)$ compared to \FsfTz, \FsfTs, \FsfTq, and \FsfTo with
$0.103(5)$, $0.104(4)$, $0.125(12)$, and $0.116(5)$, respectively.

\subsection{Supersonic simulations}
\label{sec:supersonic}
The supersonic simulations \FdTzSup, \FsfTzSup, and \FsfToSup exhibit a very similar 
behavior to their subsonic counterparts.
All simulations reach a stationary regime after $\approx$ 2.5T with the same $\Ms \approx 2.1$,
but different $\Ma$ of $2.84(13)$, $1.96(19)$, and $1.62(7)$, respectively.
Again, a higher $\Ma$ for similar $\Ms$ in the saturated regime 
implies less effective magnetic field amplification in the \FdTzSup case.
Similarly, the ratio of compressive versus rotational power in the kinetic energy spectrum
decreases from \FdTzSup to \FsfTzSup to \FsfToSup.
However, the differences are less pronounced compared to the subsonic regime given that there
is overall more power in compressive modes as expected from the supersonic regime.
The dynamical alignment between velocity and acceleration field in the supersonic regime is
virtually identical to the corresponding subsonic simulations shown in 
Fig.~\ref{fig:acc-angle}(b).

Quantitative differences between both regimes are first observed in the
$\rho$-$B$ correlation as ilustrated in Fig.~\ref{fig:obs}(d).
\FdTzSup and \FsfTzSup exhibit virtually no correlation with 
correlation coefficients of $0.04(5)$ and $-0.09(7)$ whereas \FsfToSup still shows
a weak anticorrelation with a coefficient of $-0.23(3)$. 
The logarithmic density PDFs of the supersonic simulations 
in Fig.~\ref{fig:obs}(e) follow the same trend observed in the subsonic simulations.
With decreasing power in the acceleration field the PDFs get a more pronounced negative skew.
Finally, derived line-of-sight magnetic field strength measurements are again systematically
affected as shown in Fig.~\ref{fig:obs}(f).
The derived value in the \FdTzSup simulation of $0.630(12)$ is closest the real value
of $2/3$ and the PDF is most peaked with a standard deviation of $0.229(17)$ whereas
\FsfTzSup and \FsfToSup underestimate the magnetic field strength with $0.562(19)$ and
$0.544(15)$, respectively, and generally broader PDFs with deviations of $0.39(2)$ and
$0.46(4)$, respectively.

\section{Discussion}
\label{sec:disc}

\subsection{Unrealistic large-scale $\delta$-in-time forcing}
While the idea of a $\delta$-in-time forcing is appealing on first sight due to its random, 
uncorrelated nature, we argue that it is unrealistic for two reasons.
First, no large-scale process (on some length scale $L_p$) in nature evolves instantaneously\footnote{
Small-scale processes, for example, energy injection from supernovae within a galaxy, can occur
almost instantaneously on the dynamical time of the galaxy.
However, this corresponds to small-scale forcing.
}.
For a $\delta$-in-time evolution with $T_p \rightarrow 0$, the characteristic velocity
of that process is $U_p \rightarrow \infty$.
This leads to the second, numerical argument.
A $\delta$-in-time forcing in a numerical simulation, by construction, is not resolving the 
physical timescale.
The timestep $\Delta_t$ in a simulation (of the type discussed in this Letter) is restricted so that
information locally travels no further than to adjacent cells.
Thus, with $U_p \rightarrow \infty$ the required timestep $\Delta_t \rightarrow 0$.
This restriction can never be satisfied.
Therefore, a large-scale $\delta$-in-time forcing is never numerically resolved.

\subsection{Forcing normalizations}
Similar to the autocorrelation time discussion, choosing a normalization to a constant
energy injection rate $\dot{E}$ over a constant RMS acceleration \RMSa is appealing at
first glance.
From a turbulence analysis point of view $\dot{E}$ automatically fixes the energy dissipation 
rate in the simulation.
However, 
in allowing the flow to reach a stationary state that has no realistic counterpart,
it also masks the effects of using an unresolved $\delta$-in-time forcing.

For finite autocorrelation times, the practical choice between normalizing by $\dot{E}$ 
and \RMSa is less important.
Normalizing by \RMSa for the stochastic forcing used here naturally leads to a statistical
constant energy injection rate if \Tcorr is adjusted appropriately \citep{Eswaran1988}.
In fact, both approaches can mimic realistic processes depending on the feedback mechanism.
On the one hand, normalizing by \RMSa can be seen as an external, self-consistent process that
regulates itself without significant feedback from the environment, for example, energy injection
from a jet.
On the other hand, normalizing by $\dot{E}$ can be seen as a process that depends on the interaction
with the environment, for example, cold mode AGN accretion \citep{Gaspari2012,Li2015,Meece2017}.

\subsection{Total pressure equilibrium and $\rho$-$B$ correlations}
The strong anticorrelation between the density and the magnetic field strength
observed in the \FsfTo run is consistent with the expectation of a (statistical) 
total pressure equilibrium \citep{Beck2003}
\begin{eqnarray}
B^2/2 + \pth = p_\mathrm{tot} \approx \mathrm{const.} \;.
\end{eqnarray}
In addition, the isothermal equation of state used in the simulations mandates
$\pth \propto \rho$.
Hence, to maintain a constant total pressure low density regions correspond to
regions with higher magnetic field strength and vice versa.
\citet{Yoon2016} also followed this reasoning and explained the lack of anticorrelation in
their \FdTz equivalent simulation by the forcing timescale.
$\delta$-in-time forcing evolves so fast that the system is not able to reach pressure equilibrium.
Thus, \FsfTz should also exhibit no significant anticorrelation.
However, we observe a moderate anticorrelation in that simulation, arguing that
the autocorrelation timescale alone is not sufficient to explain the results.

We argue that the increasing power in compressive modes injected by larger \RMSa (and, thus, smaller
\Tcorr) is the main driver behind a decreasing $\rho$-$B$ anticorrelation.
Given Alfv\'en's theorem in MHD, compression naturally leads to a positive correlation 
between $\rho$ and $B$. 
Any compression that is locally not exactly aligned with the magnetic field direction compresses
the magnetic field in the other two directions, which results in an increased magnetic flux.
Thus, the total pressure equilibrium induced strong $\rho$-$B$ anticorrelation is successively
weakened by increasing compressive modes associated with a positive $\rho$-$B$ correlation.
This is also in agreement with the supersonic simulations, which naturally have more power 
in compressive modes.

\subsection{Mach number dependency}
The disparity of the $\rho$-$B$ correlations for the different forcing parameters
is less pronounced with increasing sonic Mach number.
We expect that it becomes negligible in the hypersonic ($\Ms \gtrsim 5$) regime as,
for example, found in molecular clouds, because compressive modes
become dynamically important independent of the forcing scheme.

On the other hand, there is no indication that the density and LOS $B$ PDFs of 
\FdTz and \FsfTz or \FsfTo generally converge with increasing Mach number.
While we expect that the autocorrelation time (and, thus, the power in the
acceleration field) becomes less important for the statistics 
if the acceleration field
is normalized to $\left < a \right >$ (i.e., \FsfTz and \FsfTo), 
the bias for \FdTz is likely to remain.
We base this expectation on the the increasing extreme variations in 
$\left < a \right >$ in comparison to the velocity dispersion when the
acceleration field is normalized to~$\dot{E}$.

Overall this suggests that in the hypersonic regime a $\delta$-in-time forcing 
with constant power (while still being unrealistic) 
can probably be used without major implications on turbulence statistics.
Nevertheless, a more detailed study is required to verify this statement.

\subsection{Observational consequences}
An empirical relation between the derived line-of-sight magnetic field strength, the 
sonic Mach number, and the widths of the rotation measure distribution was
suggested by \citet{Wu2015}.
This relation targets rapid estimates of the LOS magnetic field in the turbulent warm ionized 
medium in our Galaxy.
However, the relation is based on simulations employing \FdTz forcing and assumes no 
$\rho$-$B$ correlation as reported by \citet{Wu2015}.
While there are differences in our \RMSa-normalized simulations, the differences are overall 
much less pronounced compared to what is observed in \FdTz.
Thus, a modified LOS $B$ estimate could still be obtained and will be explored in future work.

Similarly, magnetic field estimates in the plane of the sky as proposed by
\citet{Davis1951} and \citet{Chandrasekhar1953} are potentially affected by the
observed $\rho$-$B$ anticorrelation.
Their estimate assumes underlying isotropic Alfv\'enic perturbations of the flow.
However, the subsonic, super-Alfv\'enic regime we are probing is potentially
dominated by slow modes, which is inferred from the observed $\rho$-$B$ anticorrelation
\citep{Passot2003}.
This suggests a review of the Davis-Chandrasekhar-Fermi method for different regimes.

\subsection{Limitations}
The main purpose of the present study is to highlight and explain observed differences
in turbulence simulations employing one of the most commonly used idealized setups: 
isothermal, solenoidally driven, stationary turbulence.
Independent of the driving scheme analysis, our results indicate that compressibility and 
pressure dynamics leave a clear imprint on the flow and derived observables.
Thus, other factors are also expected to be dynamically relevant, in particular 
compressive modes in the acceleration field 
\citep{Federrath2016a}, 
and an adiabatic  equation of state \citep{Nolan2015}.

Moreover, all our simulations are super-Alfv\'enic, i.e., on average kinetic motions dominate 
magnetic field dynamics.
The super-Alfv\'enic regime could be relaxed in two directions.
On the one hand, increasing the background magnetic field strength decreases the 
Alfv\'enic mach number $\Ma$.
While \citet{Yoon2016} conducted and analyzed simulations with varying $\Ma$, the 
interplay between compressive modes and a dynamically important background field remains open.
On the other hand, we expect to see magnetic field unrelated features, for example, 
the link between
\RMSa and compressive modes, or the increasing alignment of velocity and acceleration field
with increasing \Tcorr, also in the pure hydrodynamic case.

Finally, we conducted all the simulations at a resolution of $512^3$, which is nowadays commonly
used for turbulent boxes.
While a higher resolution (and, thus, a larger dynamical range) is preferable, we observe
the same effects in higher resolution simulations at individual points in the probed 
parameter space \citep{Grete2017a}.
Thus, there is no indication that the described processes are affected by the resolution of 
the simulation.

We leave a more detailed analysis of effects pertaining to compressibility, the equation of state,
and the strength of a background magnetic field to future work.

\section{Conclusions}
\label{sec:conclusions}
In this Letter, we studied the effects of driving parameters on statistical quantities and 
observables in stationary, isothermal, compressible MHD turbulence simulations.
All simulation were driven to reach the same subsonic (supersonic) Mach number of 
$\Ms \approx 0.5$ ($\Ms \approx 2.1$) with varying autocorrelation time and normalization 
of the acceleration field.
We varied the autocorrelation time between \Tcorr$ = \{0, 1/16, 1/4,1\}$T dynamical times, i.e.,
between an effective $\delta$-in-time forcing and a forcing field that evolves on the dynamical
timescale of the flow.
In these simulations (identified with \FsfTz, \FsfTs, \FsfTq, and \FsfTo) 
the acceleration field was normalized so that the power in the acceleration field
\RMSa is constant over time where shorter \Tcorr necessitate higher \RMSa.
For the $\delta$-in-time case, we also varied to normalization in order to keep the energy 
injection rate $\dot{E}$ exactly constant over time (\FdTz).
This allowed for varying power in the acceleration field given the current flow configuration.

Our main findings are the following:
\begin{itemize}
\item With increasing \RMSa more power is injected in compressible modes even though the 
acceleration field itself is purely solenoidal.
In the most extreme case, \FdTz,  the resulting power in compressive modes becomes dynamically 
relevant so that the total kinetic energy spectrum exhibits a steeper slope compared to the other
simulations.
In addition, the relatively weaker rotational modes in \FdTz result in less efficient small-scale
dynamo action and, in turn, a lower magnetic energy saturation value.
\item With increasing \Tcorr energy injection is more efficient (and, thus, require lower \RMSa)
because the acceleration and velocity field are more aligned.
\item In the subsonic regime, there  is a strong anticorrelation 
(correlation coefficient of -0.81) 
between density and magnetic field strength for \FsfTo.
With decreasing \Tcorr (increasing \RMSa) the anticorrelation constantly weakens down to -0.5 for
\FsfTz. 
Effectively no correlation is observed for \FdTz.
The anticorrelation itself in the present regime is expected from a total pressure equilibrium.
We attribute the decreasing anticorrelation from increasing compressive modes, which are associated
with a positive $\rho$-$B$ correlation due to Alfv\'en's frozen in flux theorem.
\item In the supersonic regime, $\rho$-$B$ correlation coefficients  
are generally higher, which supports the argument for the importance of compressive modes.
\item We confirm that the presence of a $\rho$-$B$ anticorrelation leads to bias in observables, for
example, an underestimated and more variable line-of-sight magnetic field derived from rotation 
measures in the sub- and supersonic regime.
\end{itemize}

Overall, we argue that large-scale $\delta$-in-time forcing is neither realistic nor 
numerically resolved,
and conclude that results from simulations with a $\delta$-in-time forcing should 
be interpreted with care.
As such, our findings have implications for observations of magnetized turbulence in both astrophysical
and terrestrial environments, in particular those that depend on probability distribution functions
of fluid quantities \citep[e.g.,][]{Kroupp2018} or on correlations between turbulent fluctuations
\citep[e.g.,][]{Wu2015}.
A more detailed analysis of the interplay between compressive modes and pressure fluctuations 
in the context of observations will be presented in future work.

\acknowledgments
The authors thank David Collins, Alexei Kritsuk, Jeffrey Oishi, and Wolfram Schmidt for useful discussions.
PG and BWO acknowledge funding by NASA Astrophysics Theory Program
grant \#NNX15AP39G.
Sandia National Laboratories is a multimission laboratory managed and operated by 
National Technology and Engineering Solutions of Sandia LLC, a wholly owned 
subsidiary of Honeywell International Inc., for the U.S. Department of Energy's 
National Nuclear Security Administration under contract DE-NA0003525.
The views expressed in the article do not necessarily represent the views of the 
U.S. Department of Energy or the United States Government.
BWO acknowledges additional funding by NSF AAG grant \#1514700.
The simulations were run on the NASA Pleiades supercomputer through allocation SMD-16-7720 
and on the Comet supercomputer as part of the Extreme Science and Engineering Discovery Environment \citep[XSEDE][]{XSEDE}, which is supported by National Science Foundation grant number ACI-1548562, through allocation \#TG-AST090040.
\textsc{Athena} is developed by a large number of independent
researchers from numerous institutions around the world. Their
commitment to open science has helped make this work possible.

\bibliographystyle{yahapj}

\end{document}